\documentclass[letterpaper,notoc]{CECS}

\title{\emph{pp} Waves of Conformal Gravity with
Self-Interacting Source}

\author{Eloy Ay\'on-Beato$^{1,2}$\footnote{\emph{E-mail}:
\texttt{ayon-at-cecs.cl}}~~and
Mokhtar Hassa\"{\i}ne$^{1}$\footnote{\emph{E-mail}:
\texttt{hassaine-at-cecs.cl}}\\
$^{1}$Centro de Estudios Cient\'{\i}ficos (CECS),
Casilla 1469, Valdivia, Chile.\\
$^{2}$Departamento de F\'{\i}sica, Centro de Investigaci\'on y de
Estudios Avanzados del IPN,
Apdo. Postal 14-740, 07000, M\'exico D.F., M\'exico.}

\preprint{{\tiny CECS-PHY-04-14}}

\abstract{Recently, Deser, Jackiw and Pi have shown that
three-dimensional conformal gravity with a source given by a
conformally coupled scalar field admits \emph{pp} wave solutions.
In this letter, we consider this model with a self-interacting
potential preserving the conformal structure. A \emph{pp} wave
geometry is also supported by this system and, we show that this
model is equivalent to topologically massive gravity with a
cosmological constant whose value is given in terms of the
potential strength.}

\begin{document}

\section{Introduction}

In three dimensions, the matrix gauge connection
$(A_{\mu})^{\alpha}_{\beta}$ from which non-Abelian Chern-Simons
Lagrangian is constructed can be replaced by the Christoffel
connection $\Gamma^{\alpha}_{~\mu\beta}$. The resulting
Chern-Simons action reads
\begin{eqnarray}
S(\Gamma)=\frac{1}{2\kappa}\int
d^3x\,\epsilon^{\alpha\beta\gamma}\left(
\frac{1}{2}\Gamma^{\rho}_{~\alpha\sigma}\partial_{\beta}
\Gamma^{\sigma}_{~\gamma\rho}+\frac{1}{3}\Gamma^{\rho}_{~\alpha\sigma}
\Gamma^{\sigma}_{~\beta\tau}\Gamma^{\tau}_{~\gamma\rho}\right),
\label{eq:christoffelaction}
\end{eqnarray}
where $\kappa$ is a dimensionless constant. The variation of this
action with respect to the metric leads to
$$
\delta S(\Gamma)=-\frac{1}{2\kappa}\int
d^3x\,\sqrt{-g}\,C^{\mu\nu}\delta g_{\mu\nu},
$$
where $C^{\mu\nu}$ is the so-called Cotton tensor\footnote{Our
conventions are the following: the signature is $(-++)$,
$\epsilon_{012}=+1$, the Riemann tensor is
$R_{\mu\nu\rho}^{\,\;\quad\sigma}=+
\partial_{\nu}\Gamma^{\sigma}_{~\mu\rho} - \ldots$ and the Ricci tensor is
$R_{\mu\nu}=+\partial_\alpha\Gamma^\alpha_{~\mu\nu} - \ldots$.}
whose expression after using the Bianchi identity is given by
\begin{eqnarray}
C^{\mu\nu}=\frac{1}{\sqrt{-g}}\epsilon^{\mu\alpha\beta}
D_{\alpha}\left(R_{\beta}^{~\nu}-\frac{1}{4}\delta_{\beta}^{~\nu}R\right).
\label{eq:cotton}
\end{eqnarray}
Note that this tensor is symmetric, traceless and identically
conserved.

Recently, Deser, Jackiw and Pi \cite{Deser:2004wd} have considered
a three-dimensional model with a conformal scalar field for which
gravity is governed by the Cotton tensor. They found that this
system supports a \emph{pp} wave ansatz geometry where the source
also behaves like a wave. They also showed that in a special frame
their system is equivalent to topologically massive gravity
\cite{Deser:1981wh}. In this letter, we extend this work by
including a self-interacting potential which does not spoil the
conformal invariance. We show that the resulting theory still
admits \emph{pp} wave solutions for which the source does not
behave like a wave, in accordance with its self-interacting
nature. In addition, we prove the equivalence of our system with
topologically massive gravity with a cosmological constant whose
value is given in term of the potential strength.

The paper is organized as follows. We first present the action of
three-dimensional conformal gravity with source given by a
conformally coupled scalar field with a conformally invariant
self-interacting potential. Then, we solve the field equations for
a \emph{pp} wave ansatz, and we find two different classes of
solutions depending on the ratio between the coupling constants.
We establish the correspondence between this model and
topologically massive gravity with a cosmological constant.
Finally, exploiting this analogy, we derive explicit solutions of
the later theory.

\section{Three-dimensional conformal gravity with a
self-interacting potential}

We consider the three-dimensional conformal gravity with source
given by a conformally invariant scalar field with
self-interacting potential whose action reads
\newpage
\begin{eqnarray}
S(\psi, g)  & = & S(\Gamma)+I(\psi,g)\nonumber\\
            & = & S(\Gamma)-\frac{1}{2}\int d^3x\,
\sqrt{-g}\left(g^{\mu\nu}\partial_{\mu}\psi\partial_{\nu}\psi
+\frac{1}{8}R\psi^2+\lambda\psi^6\right). \label{eq:action}
\end{eqnarray}
Here the potential strength $\lambda$ is a dimensionless
constant and $R$ stands for the scalar curvature. The associated field equations are
\begin{equation}
C_{\mu\nu}  =  \kappa T_{\mu\nu}=
\kappa\left(\partial_{\mu}\psi\partial_{\nu}\psi
-\frac{1}{2}g_{\mu\nu}\left(
\partial_{\alpha}\psi\partial^{\alpha}\psi+\lambda\psi^6\right)
+\frac{1}{8}\left(g_{\mu\nu}\Box-\nabla_{\mu}\nabla_{\nu}
+G_{\mu\nu}\right)\psi^2\right),
\label{eq:fieldeqs1}
\end{equation}
\begin{equation}
\Box\psi-\frac{1}{8}R\psi-3\lambda\psi^5=0,
\label{eq:fieldeqs2}
\end{equation}
where $\Box=\nabla_{\mu}\nabla^{\mu}$. For this model, the gravity
is governed only by the Cotton tensor and the matter source,
in accordance with its conformal invariance, has a traceless
energy-momentum tensor. We now show that equations
(\ref{eq:fieldeqs1}-\ref{eq:fieldeqs2}) admit \emph{pp} wave
gravitational fields.

\subsection{The \emph{pp} wave solutions}

The plane-fronted gravitational waves with parallel rays
(\emph{pp} waves) are characterized by the existence of a
covariantly constant null vector field $k^\mu$ (see e.g.
\cite{Stephani:2003tm}). This allows to write the geometry in the
three-dimensional case as
\begin{eqnarray}
ds^2=-F(u,y)du^2-2dudv+dy^2, \label{eq:ppwave}
\end{eqnarray}
where $k^\mu\partial_\mu=\partial_v$. The covariantly constant
field is as well a Killing field and hence it is appropriate to
demand the same symmetry on the matter field, which in turn
implies that $\psi=\psi(u,y)$. In fact, the same conclusion is
reached assuming a full dependence on the scalar field, modulo the
gravitational equations, as was carefully shown in
Ref.~\cite{Deser:2004wd} in the free case. As was pointed in the
same reference, the only nonvanishing component of the Cotton
tensor (\ref{eq:cotton}) for the above metric is
$C_{uu}=\frac{1}{2}\partial^3_{yyy}F$, which implies that all the
components of the energy-momentum tensor must vanish except
$T_{uu}$. In particular, the vanishing of the $yy$-component gives
\begin{equation}\label{eq:Tyy}
T_{yy}=\frac{1}{8\sigma^3}\left[\left(\partial_y\sigma\right)^2
-4\lambda\right]=0,
\end{equation}
where we have introduced $\sigma=\psi^{-2}$ for convenience. For a
nontrivial self-interaction, this equation is solved only for a
strictly positive coupling constant ($\lambda>0$)
\begin{equation}\label{eq:sigma_yu}
\sigma(u,y)=2\sqrt{\lambda}y+f(u),
\end{equation}
where $f$ is an undetermined function of the retarded time. No other
restriction follows from the energy-momentum tensor in order to
specify the scalar field. We would like to point out that equation
(8) also solves the scalar field equation (5) and, hence
$\psi=\sigma^{-1/2}$ is the general solution for the scalar field.

Note that in the absence of self-interaction ($\lambda=0$),
Eq.~(\ref{eq:Tyy}) would imply that the scalar field just depends
arbitrarily on the retarded time $u$, Ref.~\cite{Deser:2004wd}.
This argument is no longer valid in the presence of a
self-interaction potential since this later causes the field to be
inhomogeneous on the spacelike direction orthogonal to the wave.
It remains only to solve the $uu-$component of
Eqs.~(\ref{eq:fieldeqs1}) that reads
\begin{equation}\label{eq:uu}
\frac{1}{2}\partial^3_{yyy}F= \frac{\kappa}{16}\left[
\partial_y\left(\frac{\partial_yF}{2\sqrt{\lambda}y+f}\right)
+\frac{\mathrm{d}^2\!f}{\mathrm{d}u^2}
\frac{2}{(2\sqrt{\lambda}y+f)^2}\right].
\end{equation}
The general solution of this equation for a coupling
constant $\lambda\neq(\kappa/16)^2$ is
\begin{equation}\label{eq:F(F1,F2,F3)}
F(u,y)=F_1(u)\left(2\sqrt{\lambda}y+f\right)^
{1+\kappa/(16\sqrt{\lambda})}
+F_2(u)\left(2\sqrt{\lambda}y+f\right)^2
+\frac{1}{2\lambda}\frac{\mathrm{d}^2\!f}{\mathrm{d}u^2}
\left(2\sqrt{\lambda}y+f\right)
+F_3(u),
\label{sol1}
\end{equation}
where $F_1$, $F_2$, and $F_3$ are undetermined functions of the
retarded time. In the derivation of the above result one of the
terms appearing after the second integration is a power in $y$
depending on $\lambda$. For the special coupling constant
$\lambda=(\kappa/16)^2$ this power is $-1$ and the third
integration of the related term gives a logarithmic contribution.
Hence, for $\lambda=(\kappa/16)^2$ the solution is
\begin{equation}\label{eq:F(F1,F2,F3)lam=()^2}
F(u,y)=\left[F_1(u)\ln\left(\frac{\kappa}{8}\,y+f\right)+F_2(u)\right]
\left(\frac{\kappa}{8}\,y+f\right)^2+\frac{128}{\kappa^2}
\frac{\mathrm{d}^2\!f}{\mathrm{d}u^2}
\left(\frac{\kappa}{8}\,y+f\right) +F_3(u),
\label{sol2}
\end{equation}
with $F_1$, $F_2$, and $F_3$ being again integration functions.

Since the curvature is just characterized by the only nontrivial
component of the Ricci tensor
$R_{uu}=\frac{1}{2}\partial^2_{yy}F$, any linear dependence on $y$
does not appear in the curvature which suggests that such dependence is
removable by coordinate transformations. This is a common feature
of the \emph{pp} wave gravitational fields \cite{Stephani:2003tm}.
In our case the appropriated coordinate change is
\begin{equation}\label{eq:coord1}
(u,v,y)\mapsto\left(u,
v+\frac{1}{2\sqrt{\lambda}}\frac{\mathrm{d}f}{\mathrm{d}u}y
+\frac{f}{4\lambda}\frac{\mathrm{d}f}{\mathrm{d}u}
-\int\mathrm{d}u\left[\frac{1}{8\lambda}
\left(\frac{\mathrm{d}f}{\mathrm{d}u}\right)^2
-\frac{F_3}{2}\right],y+\frac{f}{2\sqrt{\lambda}}\right),
\end{equation}
which is equivalent to put $f=0=F_3$ in the expressions
(\ref{sol1}) and (\ref{sol2}). Finally, the \emph{pp} wave
solutions of conformal gravity with a self-interacting source are
\begin{equation}\label{eq:F(F1,F2)}
ds^2=-\left[F_1(u)y^{\kappa/(16\sqrt{\lambda})-1}
+F_2(u)\right]y^2du^2-2dudv+dy^2,\qquad
\psi=\frac{1}{\sqrt{2\sqrt{\lambda}y}},
\end{equation}
for self-interaction coupling constant $\lambda\neq(\kappa/16)^2$,
and
\begin{equation}\label{eq:F(F1,F2)lam=()^2}
ds^2=-\left[F_1(u)\ln{y}+F_2(u)\right]y^2du^2-2dudv+dy^2,\qquad
\psi=\sqrt{\frac{8}{\kappa{y}}},
\end{equation}
for the value $\lambda=(\kappa/16)^2$, where in both cases we have
redefined appropriately the undetermined functions. It is
interesting to note that, contrary to the free case
\cite{Deser:2004wd}, the scalar field does not depend on retarded
time. In other words, the self-interaction breaks the wave-like
behavior of the source present in the free case.

Since the integration of Eq.~(\ref{eq:uu}) is different for
$\lambda=0$, a natural fair question is to ask whether there is a
link between our self-interacting solution at the zero $\lambda$
limit and the free one of Ref.~\cite{Deser:2004wd}. At the first
sight it seems that the self-interacting solution
(\ref{eq:F(F1,F2,F3)}) is singular at this limit. However, after
redefining the arbitrary functions of Eq.~(\ref{eq:F(F1,F2,F3)}) as
\[
F_1 = \tilde{F} f^{-(1+\kappa/16\sqrt{\lambda})},\qquad F_2 =
\frac{\alpha}{4\sqrt{\lambda}f}
     -\frac{1}{4\lambda}\frac{1}{f}\frac{\mathrm{d}^2\!f}{\mathrm{d}u^2},
     \qquad
F_3 = \beta-\frac{\alpha}{4\sqrt{\lambda}}f
     -\frac{1}{4\lambda}f\frac{\mathrm{d}^2\!f}{\mathrm{d}u^2},
\]
the free case \cite{Deser:2004wd} is exactly recovered in the zero
$\lambda$ limit, where $\tilde{F}$, $\alpha$, and $\beta$ are the
new arbitrary functions of the retarded time.\footnote{We thank R.
Jackiw for providing us these redefinitions.} A different situation
occurs when one tries to eliminate the linear dependence in $y$ in
the \emph{pp} wave geometries, since the coordinate change in the
self-interacting case (\ref{eq:coord1}) is singular as $\lambda$
goes to zero.

A study of Killing equations allows to conclude that for generic
wave profiles $F_i(u)$, $i=1,2$, the above geometries have only
the original Killing field $\partial_v$. There exists another
Killing field for the special election $F_i(u)=A_i/u^2$, $i=1,2$,
given by
\begin{equation}\label{eq:k2}
u\partial_u-v\partial_v,
\end{equation}
whose commutator with $\partial_v$ closes on $\partial_v$. The
corresponding isometry is the scaling
$(u,v,y)\mapsto(\alpha{u},\alpha^{-1}v,y)$.

\section{Correspondence to topologically massive gravity with a
cosmological constant}

We now establish the connection between the model previously
studied and the topologically massive gravity \cite{Deser:1981wh}
with a cosmological constant.

It is interesting to note that, up to a boundary term, in the
frame $g^{\prime}_{\mu\nu}=({\psi}/{\psi_0})^4g_{\mu\nu}$ the
scalar field action $I(\psi,g)$ can be seen as the
Einstein-Hilbert action with a cosmological constant, i.e.,
$$
I(\psi,g)=-\frac{\psi_0^2}{16}\int
\left[R(g^{\prime})-2(-4\lambda\psi_0^4)\right]\sqrt{-g^{\prime}}\,d^3x.
$$
Here, $\psi_0^2$ is a constant with dimension of the inverse of
the length in order to have dimensionless actions in both sides of
this relation. For convenience, we choose this constant to be $
\psi_0^2=1/(\pi G)$ where $G$ is the Newton's constant. Moreover,
since the Chern-Simons action (\ref{eq:christoffelaction}) is
conformally invariant, the full action $S$ defined in
(\ref{eq:action}) becomes in the new frame
\begin{eqnarray}
S=S(\Gamma)-\frac{1}{16\pi G}\int
\left(R-2\Lambda\right)\sqrt{-g}\,d^3x, \label{newaction}
\end{eqnarray}
where we have dropped the primes and $\Lambda$ is the cosmological
constant given in terms of the potential strength by
\begin{eqnarray}
\Lambda=-\frac{4\lambda}{\pi^2 G^2}. \label{lambda}
\end{eqnarray}
The field equation associated to (\ref{newaction}) reproduce those
of topologically massive gravity with a cosmological constant,
i.e.,
\begin{eqnarray}
\frac{1}{\mu}C_{\mu\nu}+G_{\mu\nu}+\Lambda g_{\mu\nu}=0,
\label{neweqsdim}
\end{eqnarray}
where the topological mass is expressed as
$\displaystyle{\mu=-\frac{\kappa}{8\pi G}}$. Note that, in order
to make contact with topologically massive gravity the signs of
the topological mass and the dimensionless parameter $\kappa$ must
be different. This ambiguity is irrelevant when the constants of
the problem are not tied, since we are just mapping a solution of
Eqs.~(\ref{eq:fieldeqs1}-\ref{eq:fieldeqs2}) to one of
Eqs.~(\ref{neweqsdim}). Indeed, using the correspondence together
with the \emph{pp} wave solutions (\ref{eq:F(F1,F2)}) found
previously for $\lambda>0$ and $\lambda\neq(\kappa/16)^2$, it is
straightforward to see that the following spacetimes
\begin{equation}\label{eq:F(F1,F2)YF}
ds^2=l^2\left\{-\left[F_1(u)\mathrm{e}^{-(\mu{l}+1)y}+F_2(u)\right]du^2
-2\mathrm{e}^{-2y}dudv+dy^2\right\},
\end{equation}
are solutions of topologically massive gravity with a negative
cosmological constant $\Lambda=-1/l^2$. As anticipated, the sign
of the topological mass is irrelevant in this case and can be
chosen to be positive. This last remark is no longer valid for the
other \emph{pp} wave solution (\ref{eq:F(F1,F2)lam=()^2}) since
the potential strength $\lambda$ and the dimensionless constant
$\kappa$ are not independent constants. Indeed, the correspondence
to topologically massive gravity with a negative cosmological
constant exists provided the topological mass to be negative and
given by $\mu=-1/l$. In this case, the spacetime geometry given by
\begin{equation}\label{eq:F(F1,F2)lam=()^2YF}
ds^2=l^2\left\{-\left[F_1(u)y+F_2(u)\right]du^2
-2\mathrm{e}^{-2y}dudv+dy^2\right\},
\end{equation}
is solution of equations (\ref{neweqsdim}) with a negative
cosmological constant $\Lambda=-1/l^2$ and a negative topological
mass.

To end with this section, we would like to stress that the
solutions of topologically massive gravity obtained through the
correspondence are no longer \emph{pp} waves, since these
spacetimes do not admit covariantly constant null fields.

\section{Discussion}

Here, we have shown that $(2+1)-$dimensional conformal gravity
with source given by a conformally invariant self-interacting
scalar field admits \emph{pp} wave solutions. The corresponding
scalar sources do not behave like a wave, since they are
independent of the retarded time and inhomogeneous in the
spacelike direction orthogonal to the wave. Obviously, this is a
manifestation of its self-interacting character. We found just one
particular wave profile supporting an additional isometry of the
\emph{pp} waves, which corresponds to a special scaling of the
retarded time and the null coordinate. We have established a
correspondence between this model and topologically massive
gravity with a cosmological constant. Finally, using this analogy
we have obtained explicit solutions of the later theory for a
negative cosmological constant. It would be interesting to explore
whether the conformal gravity supports other physical interesting
configurations.

\bigskip

\bigskip

\noindent\emph{Acknowledgments.-} We thank C. Mart\'{\i}nez, C.
N\'u\~{n}ez, R. Troncoso, and J. Zanelli for discussions and
specially to S. Deser and R. Jackiw for enlightening comments. This
work is partially supported by grants 3020032 and 1040921 from
FONDECYT, grants 38495E and 34222E from CONACyT, grant 2001-5-02-159
from CONICYT/CONACyT, and grant D-13775 from Fundaci\'on Andes.
Institutional support to the Centro de Estudios Cient\'{\i}ficos
(CECS) from Empresas CMPC is gratefully acknowledged. CECS is a
Millennium Science Institute and is funded in part by grants from
Fundaci\'{o}n Andes and the Tinker Foundation.



\begin{thebibliography}{99}

\bibitem{Deser:2004wd}
S.~Deser, R.~Jackiw and S.~Y.~Pi,
``Cotton blend gravity pp waves,'' arXiv:gr-qc/0409011.

\bibitem{Deser:1981wh}
S.~Deser, R.~Jackiw and S.~Templeton,
Annals Phys. \textbf{140}, 372 (1982)
[Erratum-ibid. \textbf{185}, 406 (1988)];
Phys.\ Rev.\ Lett.\ \textbf{48}, 975 (1982).

\bibitem{Stephani:2003tm} H.~Stephani, D.~Kramer, M.~MacCallum,
C.~Hoenselaers and E.~Herlt,
\emph{Exact solutions of Einstein's field equations}
(Cambridge University Press, Cambridge 2003).

\end{thebibliography}
\end{document}